# An Automated Validation Framework for Power Management and Data Retention Logic Kits of Standard Cell Library

Akshay Kamath, Bharath Kumar, Sunil Aggarwal, Subramanian Parameswaran,
Parag Lonkar, Debi Prasanna, Somasunder Sreenath
Samsung Semiconductor India R&D Center
Bengaluru, India
{akshay.k2, bharath.k1, sunil.agg, subbu.param, parag.lonkar, debi.das, soma.ks}@samsung.com

*Abstract*-The development of a standard cell library involves characterization of a number of gate-level circuits at various cell-level abstractions. Verifying the behavior of these cells largely depends on the manual skills of the circuit designers. Especially challenging are the power management and data retention cells which must be checked thoroughly for voltage and power configurations in addition to their logic functionality. Also, when standard cells are extracted into various models, any inconsistencies in these models typically goes unchecked during library development. Thus, validating these cells exhaustively prior to customer delivery is highly advantageous to not only improve customer satisfaction but also to reduce design costs. We address this challenge by presenting a methodology to validate the power management and data retention cells that are used in the logical design flow of low-power chips. For a quick adoption by standard cell library design teams, the framework is fully automated and runs out-of-the-box. The proposed framework has been implemented and deployed within the Samsung Foundry ecosystem to enhance the overall quality of library design kit deliverables.

## I. INTRODUCTION

Standard cell libraries make up an essential component of Samsung Foundry's ecosystem for chip development. To deliver a library design kit of highest quality, the standard cells should be thoroughly validated in terms of standalone functionality verification as well as their system-level use-case correctness. In an IP or system-on-chip design flow where multiple library definitions are in use (e.g. Liberty model for synthesis and Verilog model for simulations), identifying any inconsistencies present in these definitions is of vital importance. To do so, a robust system-level testing framework is required during the initial stages of standard cell library development.

Validation of power management and data retention logic kits, which comprise of isolation cells, level shifters, enable level shifters, and data retention cells, is particularly challenging. This is because the power-intent strategies required for inference of these cells in a design, usually specified through the industry standard Unified Power Format (UPF) [1], vary considerably amongst the various cell types. For instance, the isolation could be clamp-low or clamp-high, level shifters could be to step-up or step-down or both, retention cells could be flip-flop based or latch based, etc. In addition, a cell type can have sub-types depending on drive strengths, connection to local or global power rails, source or sink etc. Thus, in order to set up a validation framework for these cells, we need an infrastructure that mimics the usage of standard cells in a system-level environment that comprises of voltage islands and gated power domains. For this purpose, we make use of a specific variation of the generic multi-voltage design platform [2] that we previously developed. Given such an environment which satisfies the power and voltage conditions required to infer each of the power management and data retention cells, we develop a framework to –

- Verify if the relevant cells get inferred with appropriate power pin connections at the intended locations
- Perform simulation-based verification of cells including their power dependencies and DFT compatibilities
- Automate the entire process of generating the netlists as well as the verification test cases

The design infrastructure is explained in Section II along with the techniques for inferring the power management and data retention cells. The simulation-based verification setup is described in Section III and the automation flow in Section IV. The effectiveness of the proposed framework in capturing various issues is illustrated in Section V and conclusions are drawn in Section VI.

## II. DESIGN INFRASTRUCTURE

The design microarchitecture comprises of a set of register blocks, depicted in Fig. 1, which can be configured to operate at designated voltage levels as always-on power domains or switchable power domains. Each of these blocks defines a number of unique address-mapped registers, known as Special Function Registers or SFRs. These registers are made accessible over the AMBA Advanced Peripheral Bus (APB) [3] with each block implementing the protocol's handshake logic. The motivation for choosing this standard bus-based architecture has to do with the ease of generating these register blocks automatically – using a front-end EDA software such as Magillem [4] – while being a reasonable representation of usage of power management and data retention cells in a system-level environment. Though other standard protocols (such as AHB or AXI) or even a custom protocol can be used with same effectiveness, we employ the APB protocol for its minimalist design that helps achieve our purpose.

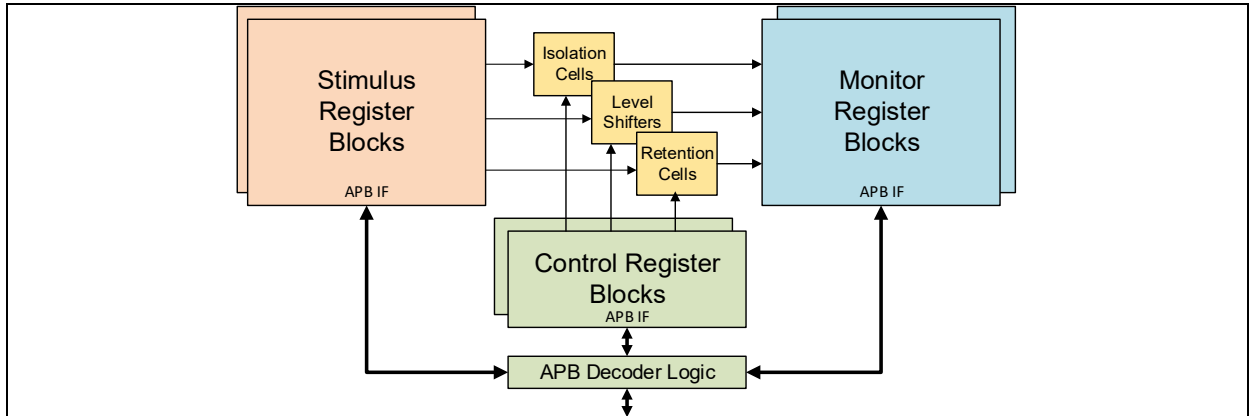

Fig. 1: System-level design platform for validating standard cells

The stimulus blocks comprise of read-write SFRs for driving stimuli on the input data pins of standard cells while the monitor blocks include read-only SFRs to observe their output responses. A control SFR block contains registers to drive control pins of applicable cells such as enables and resets. Each of these blocks is implemented as a separate RTL module. The RTL does not include any instance of standard cell and is not power-aware. The power information for inferring the cells are provided using UPF constraints.

We operate at two voltage $V_H$ V and $V_L$ V where $V_H > V_L > 0$. The ground pin "VSS" is taken as 0 V for all blocks. Global power rails are indicated as "VDDG1" and "VDDG2" while power switches are indicated by "SW". The following sub-sections illustrate the operating conditions required to infer all the power management and data retention cells between pairs of stimulus and monitor blocks.

*A. Data Retention Cells*

Data retention logic kit includes special flip-flops and latches which can store their states when the block in which they are present is powered down and also restore their states when the power is back up. These cells, therefore, can only be inferred in a gated power domain. Since the register blocks are auto-generated, we do not tamper with those blocks for inferring the retention cells. Instead, we create a new module called retention island (RET_ISLAND) between a pair of stimulus and monitor SFR blocks as shown in Fig. 2. Within this island module, different kinds of flip-flops and latches are behaviorally modeled and UPF constraints are developed to replace these with their retention counterparts after synthesis.

*B. Isolation Cells*

Isolation cells are used to detach signals crossing from a power-gated domain to an always-on domain. Thus, using a power-gated stimulus SFR block and an always-on monitor SFR block, both operating at same voltage levels ($V_H$), isolation cells can be inferred at the input or the output boundaries of the stimulus SFR block as shown in Fig. 3. Depending on how the input signal is isolated, we typically have AND-type, OR-type and Latch-type isolation cells. Cells "ISO1" and "ISO2" indicate the different types of isolations cells with and without backup power pins respectively. While the former ones are inferred inside the power-gated block, the latter ones are inferred outside in the always-on top-level wrapper.

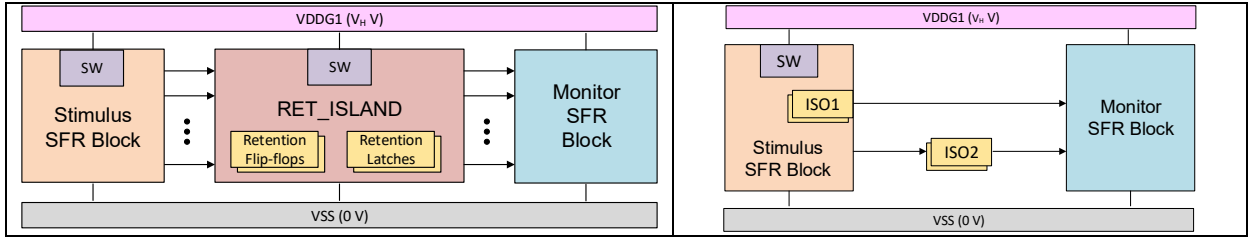

Fig. 2: Operating conditions required to infer data retention cells

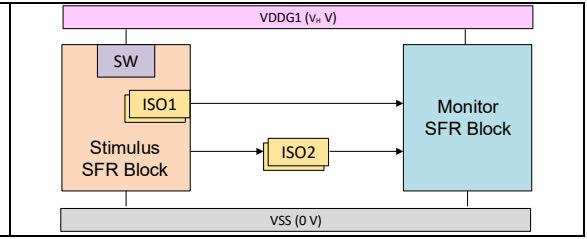

Fig. 3: Operating conditions required to infer isolation cells

*C. Level Shifters*

When a signal crosses from one always-on domain to another that are operating at different voltages, a level shifter is required to step-up or step-down the signal's voltage level. Using an always-on stimulus SFR block at $V_H$ V and two monitor blocks at $V_L$ (with one being always-on and the other being power-gated), level shifters can inferred at the input or the output boundaries of the monitor blocks as shown in Fig. 4. Cells "LS1" indicate sink-type up-down level shifters without backup power pins while "LS2" are their source-type counterparts. "LS3" are sink-type up-down level shifters with backup power pins and are thus inferred in a power-gated monitor block.

*D. Enable Level Shifters*

Enable level shifters are connected between power-gated domains and always-on domains operating at different voltage levels. These cells can conceptually be thought of as integrated cells that perform isolation as well as level shifts. The kind of stimulus and monitor SFR blocks required to infer these cells are shown in Fig. 5. Cells "ELS1" are sink-type up-down enable level shifters while "ELS2" are their source-type counterparts. "ELS3" are sink-type step-down enable level shifters. "ELS4" are sink-type up-down enable level shifters, the only ones that include backup power pins.

*E. Overall Design*

Using the state-of-the-art standard cell library of Samsung Foundry that we developed the proposed validation framework for, we were able to infer a unique instance of each of the cells in the power management and data retention logic kits for a **100%** logical cell coverage in a single netlist. The overall design is implemented using one always-on control SFR block operating at $V_H$ V, two power gated stimulus blocks at different voltage levels, and three monitor SFR blocks (two always-on blocks at either voltage levels and one power gated block operating at $V_L$ V). This is possible through the usage of secondary instances of isolation cells for level shifters. The top-level wrapper is an always-on domain operating at $V_H$ V.

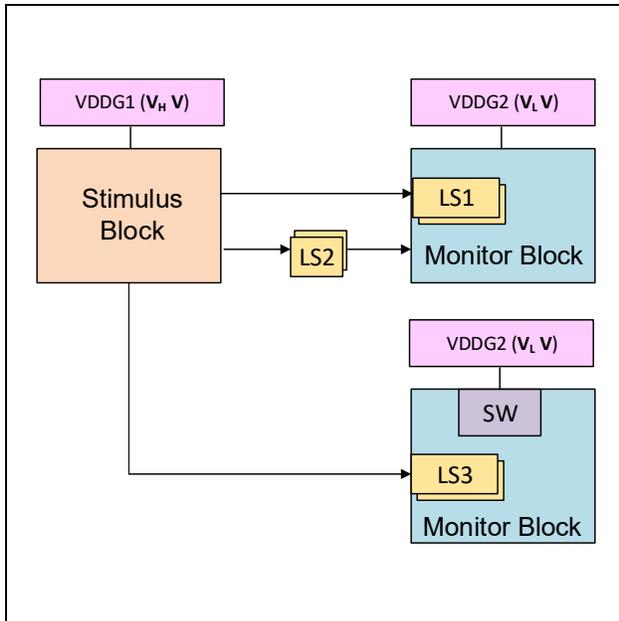

Fig. 4: Operating conditions required to infer level shifters

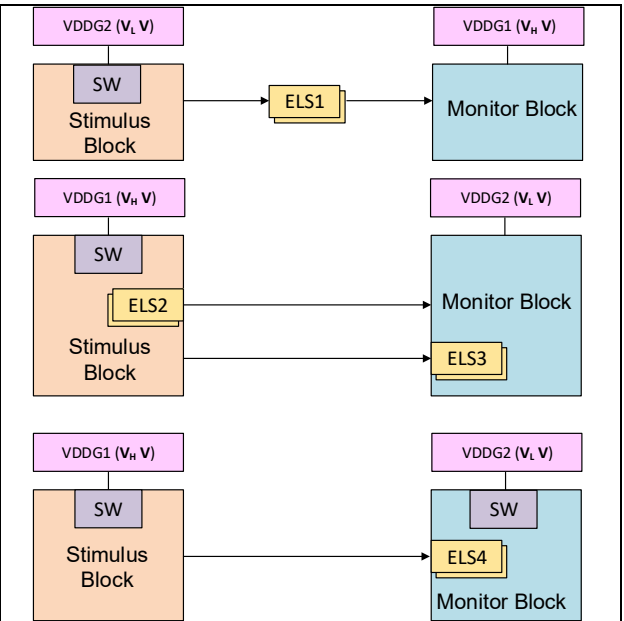

Fig. 5: Operating conditions required to infer enable level shifters

## III. SIMULATION-BASED VERIFICATION

With the power-aware gate-level netlist as the Design-Under-Test (DUT), different scenarios and checkers are created by driving desired cells with appropriate stimulus vectors and monitoring their output responses through sequences of APB write-read transactions on the address-mapped registers. While cyclic stimuli are generated for combinatorial cells, random stimuli (that change as per the constraints of corner cases) are created for sequential cells. The test bench environment for verifying the functionality of the inferred cells includes multiple checkers viz. hierarchy-based truth table check, SFR-based truth table check and hierarchy-based assertions (using cell output predictor). The test bench architecture is shown in Fig. 6.

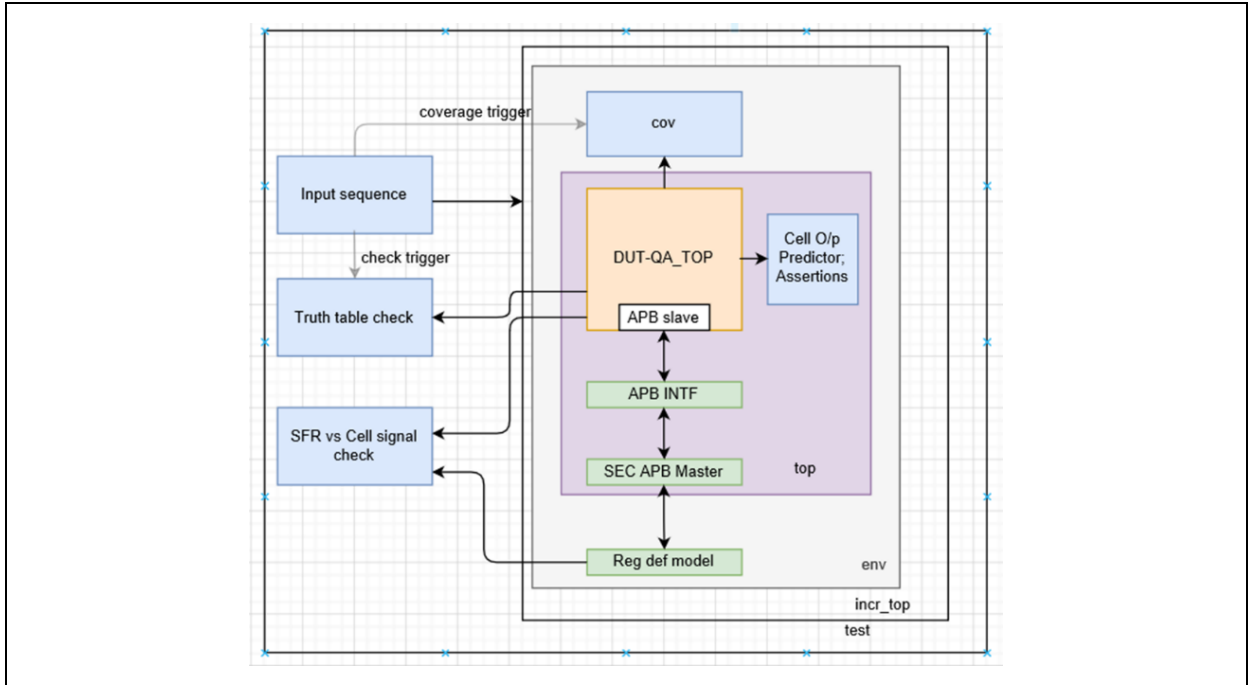

Fig. 6: Test bench architecture for verification of standard cells

The UVM register model provides a way of tracking the register content of a DUT and a convenience layer for accessing register and memory locations within the DUT. This register model is used to create stimulus using sequences. The model contains a number of access methods which sequences use to read from and write to the SFRs. These methods cause generic register transactions to be converted into transactions on the target bus. The UVM package contains a library of built-in test sequences which can be used for most of the basic register tests, such as checking register reset values and checking the register data paths.

While we can generate different kinds of stimulus for combinatorial cells, for simplicity, we use cyclic stimulus for covering all possible combinations of inputs one after the other. In case of sequential cells, using cyclic stimulus may not cover all the use-case scenarios since the current output is dependent on the previous output value. So we simplify the code by generating random stimulus vectors for sequential cells and also make the code generic using the UVM features. For instance, Fig. 7 shows the SV-UVM code snippet for cyclic stimulus used for verifying the non-power dependent functionality of combinational cells.

Functional and code coverage are used as metrics for evaluating the scenarios tested. Functional coverage is an important feature in the setup that marks the closure of the verification process when the number reaches 100%. A functional coverage monitor analysis component contains one or more cover-groups which are used to gather information related to the events that have occurred in a test-bench during a test case. The code generation of the cover-groups and cover-points of functional coverage is automated using scripts which makes the test-bench scalable for any number of cells in the design. The coverage of complex scenarios can be tracked by having transitional coverage for the same in the functional coverage. Fig. 8 shows an example of the functional coverage of a basic level shifter cell.

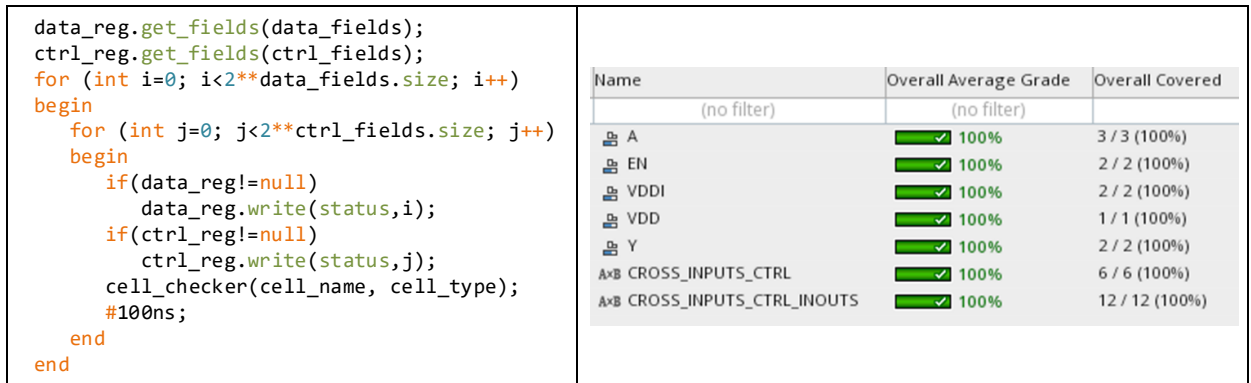

Fig. 7: UVM code snippet for cyclic stimulus generation      Fig. 8: Coverage report of an enable level shifter

The voltage supplies to the DUT can also be controlled from the test environment which are essential for testing complex scenarios of power management and data retention cells such as data retention feature, reset feature, set feature, power independent isolation feature, unknown state of cell inputs, and so on. The progress is then tracked in the cross coverage of the functional coverage. Coverage of the power pins can be covered by controlling the power switch cell which in turn controls the input power pin of the cell.

## IV. AUTOMATION FLOW

The proposed framework, as depicted in Fig. 9, involves the automatic generation of a power-aware gate-level netlist comprising of unique instances of all the power management and data retention cells. Using this netlist, a suite of test cases for simulation-based verification and a scan-inserted netlist for ATPG simulation are also automatically generated. The results of these simulations help in identifying any issues in the different standard cell library models.

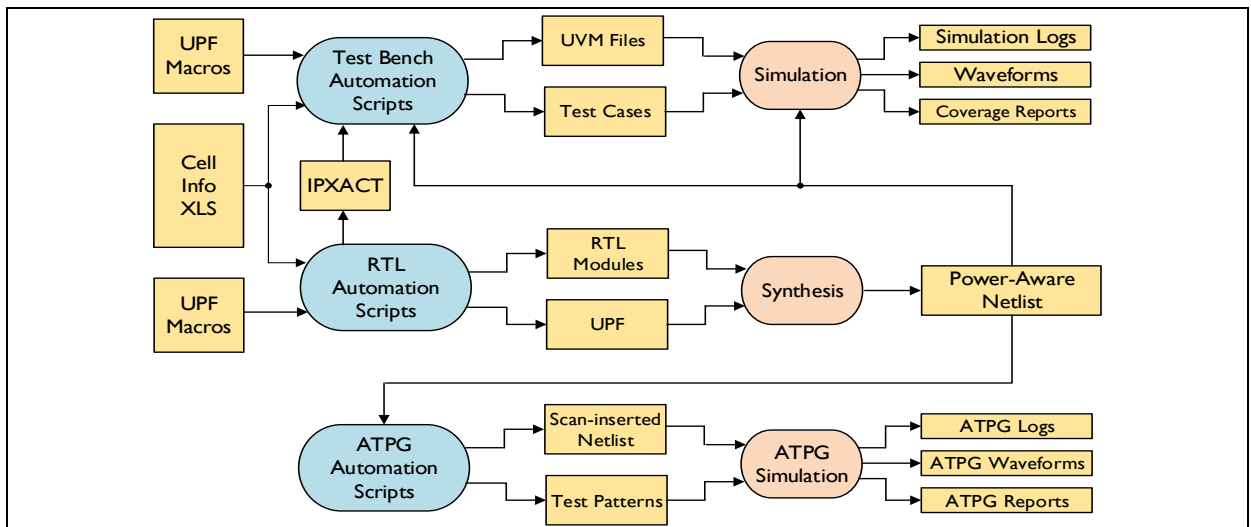

Fig. 9: RTL, Test Bench and ATPG Automation Flow Diagram

### A. Cell Info XLS

The input information to automation scripts is provided through a spreadsheet, referred to as Cell Info XLS, which contains various attributes of standard cells in a specific pre-defined format. These include, but are not limited to, the names of standard cells, their drive strengths, their logical and power ports, their type (isolation, level shifters etc.) and their functionality. The Cell Info XLS is derived from the standard cell library datasheet. Some of the parameters present in the Cell Info XLS have been illustrated in Table 1 for the following sample cells –

- Sink-side latch-type isolation cell with active high enable and a non-inverted output.
- Source-side up/down level shifter cell with active low enable.
- Negative-edge triggered static D-type retention flip-flop with scan enable, asynchronous active-high "RESET", and asynchronous active-low "SET" that dominates with a non-inverted output.

TABLE 1: Cell Info XLS Template

| Cell Index | Cell Type | Cell Inputs | | | Cell Outputs | Cell Supply Pins | Cell Attributes | |
|---|---|---|---|---|---|---|---|---|
| | | Data Signals | Control Signals | Interface Signals | Data Signals | | Cell Function | Cell Location |
| **CELL_NUM** | **CELL_TYPE** | **CELL_DATA_PINS** | **CELL_CTRL_PINS** | **CELL_IN_IF_PINS** | **CELL_MON_PINS** | **CELL_INOUTS** | **CELL_FUNCTION** | **CELL_LOCATION** |
| 1 | ISO | D | G | | Q | VDD | LATCH | parent |
| | | | | | | VSS | | |
| | | | | | | VNW | | |
| | | | | | | VPW | | |
| 2 | ELS | A | ENB | | Y | VDD | OR | self |
| | | | | | | VDDO | | |
| | | | | | | VSS | | |
| | | | | | | VPW | | |
| | | | | | | BIASNW | | |
| 3 | DRET | D | RETN | CKN | Q | VDD | FLOP | - |
| | | | R | SI | | VDDG | | |
| | | | SN | SE | | VSS | | |
| | | | | | | VNW | | |
| | | | | | | VPW | | |

## B. RTL-to-Netlist Automation Flow

Custom scripts, written in Python, extract relevant cell data from the Cell Info XLS to generate various register and island blocks as wells as the top-level wrapper with APB decoding logic. The scripts are designed to assign unique names to each cell which are used as prefixes for cell ports. The scripts also generate a spreadsheet with the appropriate register mapping information which is then converted into an IPXACT XML [5].

In addition to RTL and IPXACT, top-level UPF constraints for synthesis are created by labelling cells with common power strategies and developing UPF macros with placeholders. The UPF macros for isolation and retention cells are shown in Fig. 10 and Fig. 11 respectively. The scripts substitute the appropriate values for the corresponding cells by matching the values under the respective placeholder columns of the Cell Info XLS. The synthesis tool uses these inputs to generate a power-aware gate-level netlist that infers all the power management and data retention cells along with the necessary power port connections.

```
set_isolation CELL_"CELL_NUM" \
-domain "SUB_DOMAIN" \
-isolation_power_net "AON_POWER" \
-isolation_ground_net "PRIMARY_GROUND" \
-clamp_value "CELL_CLAMP_VAL" \
-elements {u_"RTL_INSTANCE"/ \
o_sfr_"CELL_MODULE"_data_"CELL_DATA_PINS[0]"}

set_isolation_control CELL_"CELL_NUM" \
-domain "SUB_DOMAIN" \
-isolation_signal "CONTROL_BLOCK"/ \
o_sfr_"CELL_MODULE"_ctrl_"CELL_CTRL_PINS[0]" \
-isolation_sense "CELL_SENSE" \
-location "CELL_LOCATION"

map_isolation_cell CELL_"CELL_NUM" \
-domain "SUB_DOMAIN" \
-lib_cells {"CELL_MODULE"}
```

```
set_retention_cell CELL_"CELL_NUM" \
-domain "SUB_DOMAIN" \
-retention_power_net "AON_POWER" \
-retention_ground_net "PRIMARY_GROUND" \
-elements {u_"RTL_INSTANCE"/ \
o_"CELL_MODULE"_"CELL_OUTPUT_PINS[0]"}

set_retention_control CELL_"CELL_NUM" \
-domain "SUB_DOMAIN" \
-save_signal {"CONTROL_BLOCK"/ \
i_"CELL_MODULE"_ctrl_"CELL_CTRL_PINS[0]" high} \
-restore_signal {"CONTROL_BLOCK"/ \
i_"CELL_MODULE"_ctrl_"CELL_CTRL_PINS[0]" low} \

map_retentionn_cell CELL_"CELL_NUM" \
-domain "SUB_DOMAIN" \
-lib_cells {"CELL_MODULE"}
```

Fig. 10: UPF Macro for Isolation Cells

Fig. 11: UPF Macro for Retention Cells

*C. Test Bench Automation Flow*

The first step in creating test benches for simulation-based verification involves extracting the hierarchies of all the cells present in the input netlist using a custom Python script. The next step is to generate the register-definition from the input IPXACT XML which is accomplished via a shell script (vbuilder). Another custom Python script dumps various UVM environment files for stimulus, checkers, coverage, etc. A truth table XLS, derived from the same standard cell library datasheet, is utilized to generate truth table checker files and test cases for verifying functional behavior of all the power management and data retention cells.

*D. Scan-insertion and ATPG Automation Flow*

TCL scripts are developed with DFT configuration settings for the previously generated power-aware netlist. After reading the design, the test protocol is created and DFT design rule checks are performed. Using the specified scan architecture, scan chains are inserted in the design (scan compression is disabled). ATPG model is then built for the new scan-inserted netlist and the corresponding test patterns (STIL, WGL, BIN) are generated for ATPG simulations.

## V. FRAMEWORK EFFICACY

In our implemented design with Samsung Foundry's state-of-the-art standard cell library, we have been able to infer at least one instance of each of the 22 unique power management cells and 18 unique data retention cells and achieved a 100% coverage result. The development of verification environment and automation scripts took a one-time effort of **4 weeks**. The scripts developed to generate the intermediate RTL modules, test benches and UPF now allow for netlist generation in under **10 minutes**. The nature of design and automation also makes the platform technology independent as the same infrastructure can be re-used for a standard cell library of a different technology node.

*A. Validation of Timing Model (.lib, .db)*

The script developed for extracting cell hierarchies also reports if any cells present in the Cell Info XLS have not been inferred in the netlist. The UPF macros for those cells can be cross-checked and modified if required. In case the UPF settings are accurate with a missing cell's use-case definition, the Liberty database can then be debugged to find the root-cause for absence of that particular cell in the netlist at the intended location.

*B. Validation of Behavioral Model (.v)*

The collection of test cases generated allow for comprehensive testing of all the power management and data retention cells in terms of logical behavior as well as power functionality. For instance, consider an AND-type global powered isolation cell with active-low enable defined as per the truth table in Table 2. This cell gets inferred inside a power-gated stimulus SFR block in the design and thus its input data pin "A" is sensitive to the block's power "VDD". The enable pin "ENB" and the output pin "Y" are sensitive to the always-on global power "VDDG" as these are driven and monitored in their respective always-on blocks. As seen in Fig. 12, the cell behaves like a bubbled AND-gate when the stimulus block is powered on.

Table 2: Truth table of an AND-type global powered isolation cell with active low enable

| Input Pin<br>A | Enable Pin<br>ENB | Switch Power Pin<br>VDD | Ground Pin<br>VSS | Always-on Power Pin<br>VDDG | Output Pin<br>Y |
|---|---|---|---|---|---|
| 0 | X | X | 0 | 1 | 0 |
| X | 1 | X | 0 | 1 | 0 |
| 1 | 0 | X | 0 | 1 | 1 |

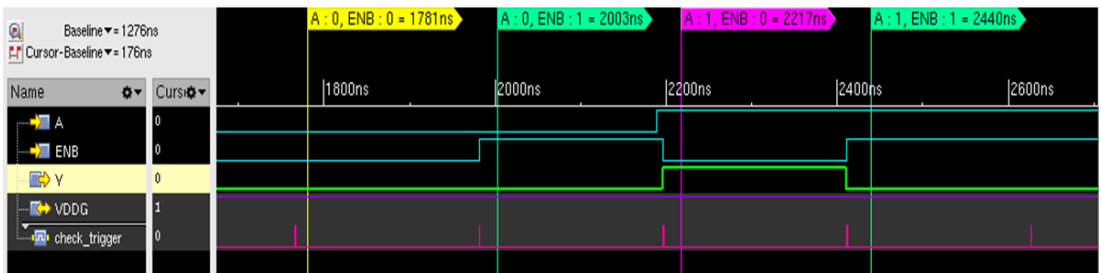

Fig. 12: Waveform snapshot of isolation cell behavior when the block is powered on

When the block's power is turned off, "A" goes to unknown state and the cell clamps the output "Y" to low if the "ENB" pin is high as shown in Fig. 13. All of these pins are driven and monitored through APB read-write transactions at different intervals of time.

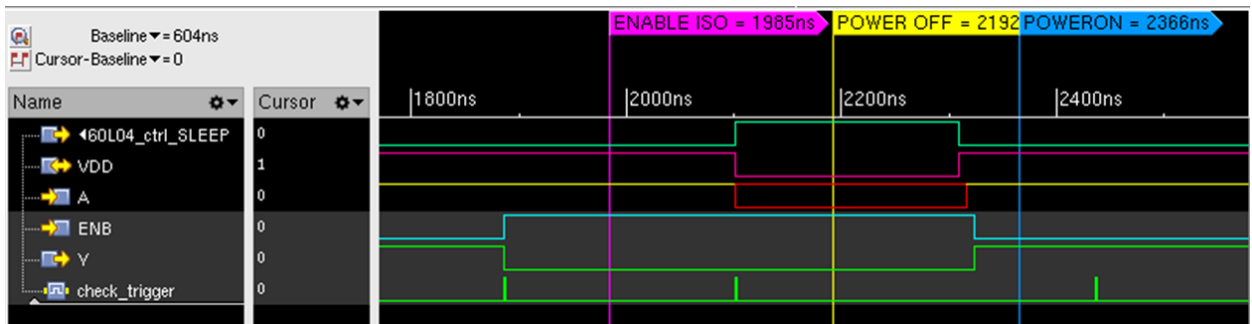

Fig. 13: Validation of isolation cell behavior when the block is powered off with isolation enabled

In addition to the above functionality checks, the proposed framework is also useful in capturing any inconsistencies in the Verilog models of the standard cell library which is otherwise not possible at cell-level views. To illustrate this feature, consider an enable level shifter defined as per the truth table in Fig. 14. Its correct Verilog model definition is shown in Fig. 15. Suppose that the actual Verilog model inadvertently has the following assignment – assign Y = ((VDDO === 1'b1) && (VDD === 1'b1) && (VSS === 1'b0))? out_temp : 1'bx;

| Input Pin A | Enable Pin EN | Switch Power Pin VDD | Ground Pin VSS | Always-on Power Pin VDDO | Output Pin Y |
|---|---|---|---|---|---|
| 0 | X | 1 | 0 | 1 | 0 |
| X | 0 | X | X | 1 | 0 |
| 1 | 1 | 1 | 0 | 1 | 1 |

Fig. 14: Truth table of an enable level shifter

```
...
and I0(out_temp, A, EN);
assign Y = ((VDDO === 1'b1)
            && (!EN|VDD === 1'b1)
            && (VSS === 1'b0))?
            out_temp : 1'bx;
...
```

Fig. 15: Correct Verilog model of the enable level shifter

With the aid of checkers and assertions generated from the truth table, such an incompatible Verilog model definition can be detected in simulations as illustrated in Fig. 16.

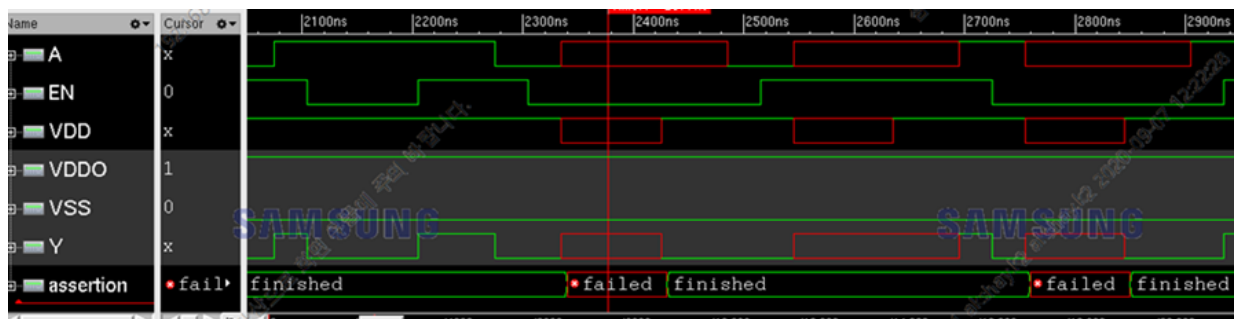

Fig. 16: Simulation waveform snapshot of the enable level shifter

The validation framework can address other similar issues in standard cell library development. For instance, when the Verilog model of any power management cell fails to include the dependency of the cell's output signal on the state of a power-gated domain, driving the cell's inputs to known values and switching off the corresponding power domain can capture any inconsistency that may exist in the descriptions of power-down function between Verilog and Liberty views for such cells. Similarly, for data retention cells which follow a definite sequence for save and restore, simulations with error-sequences help in capturing any differences between the Verilog and schematic views.

*C. Validation of ATPG Model (.tv, .mdt)*

The scan-inserted netlist is used to run ATPG simulation using the test patterns generated. Stuck-at fault coverage reports are generated that can be analyzed for detectable and undetectable faults. Scan DRCs run on the scan-inserted netlist check if the scan registers can be controlled and if the scan chains are traced properly or not. Testability analysis can help capture the controllability and observability issues from testing perspective.

This flow also allows to capture inconsistencies between the Verilog model and ATPG model definitions of retention cells. For instance, in one of our previous libraries, it was observed at a much later stage that the retention mode for retention cells was modelled differently in the Verilog and ATPG models. While the Verilog model behavior for emulating the output changes when the cells were clocked in retention mode was exactly the same as the circuit behavior, this aspect was skipped in the ATPG model. Such issues can now be easily caught in ATPG simulations using the generated scan-inserted netlist and the ATPG models can be updated to include the missing features.

## VI. Conclusion

The validation framework introduced in this paper for power management and data retention library cells allows for detection (and thereby correction) of system-level bugs prior to database delivery, which might otherwise be caught at later stages leading to quality concerns and increased support requirement. The infrastructure for implementing this framework has been developed and successfully tested on Samsung Foundry's standard cell libraries.